\documentclass[twocolumn,aps,floats,letterpaper,floatfix,groupedaddress,eqnsec]{revtex4-1}
\usepackage{graphicx}
\usepackage{dcolumn}
\usepackage{amsmath}
\usepackage{amsfonts}
\usepackage{amssymb}
\usepackage{epsfig,float,afterpage,wrapfig,psfrag}
\usepackage{natbib}

\newcommand{\beq}{\begin{equation}}
\newcommand{\eeq}{\end{equation}}
\newcommand{\bes}{\begin{subequations}}
\newcommand{\ees}{\end{subequations}}
\newcommand{\bea}{\begin{eqnarray}}
\newcommand{\eea}{\end{eqnarray}}
\newcommand{\ba}{\begin{array}}
\newcommand{\ea}{\end{array}}
\newcommand{\beqn}{\begin{eqnarray*}}
\newcommand{\eeqn}{\end{eqnarray*}}

\newcommand{\f}[2]{\frac{#1}{#2}}

\newcommand{\la}{\langle}
\newcommand{\ra}{\rangle}

\def\nn{\nonumber}

\newlength{\sizeonefig}
\newlength{\sizetwofig}
\begin{document}

\title{Spin noise spectroscopy of quantum dot molecules}

\author{Dibyendu Roy$^{1,2}$, Yan Li$^3$, Alex Greilich$^4$, Yuriy V. Pershin$^5$,  Avadh Saxena$^1$, Nikolai A. Sinitsyn$^1$}

\affiliation{$^1$Theoretical Division, Los Alamos National
Laboratory, Los Alamos, New Mexico 87545, USA}

\affiliation{$^2$Center for Nonlinear Studies, Los Alamos National Laboratory, Los Alamos, New Mexico 87545, USA}

\affiliation{$^3$National High Magnetic Field Lab, Los Alamos
National Laboratory, Los Alamos, New Mexico 87545, USA}

\affiliation{$^4$Experimentelle Physik 2, Technische Universit\"at
Dortmund, D-44221 Dortmund, Germany}

\affiliation{$^5$Department of Physics and Astronomy and USC
Nanocenter, University of South Carolina, Columbia, SC 29208, USA}

\pacs{72.70.+m, 72.25.Rb, 78.67.Hc} 

\begin{abstract}

We discuss advantages and limitations of the spin noise
spectroscopy for characterization of interacting quantum dot
systems on specific examples of individual singly and doubly charged
quantum dot molecules (QDMs). It is shown that all the relevant
parameters of the QDMs including tunneling amplitudes with
spin-conserving and spin-non-conserving interactions, decoherence
rates, Coulomb repulsions, anisotropic g-factors and the distance
between the dots can be determined by measuring properties of the
spin noise power spectrum.
\end{abstract}
\vspace{0.0cm}

\maketitle
\section{Introduction}

Spins of electrons or holes in two interacting neighboring quantum
dots (QDs) represent a building block for realization of a scalable
solid-state quantum information platform~\cite{Loss98}. Such a block
unit, called a quantum dot molecule (QDM), can be e.g. electrostatically defined \cite{echo} or it
can be formed from
two closely spaced self-assembled InAs/GaAs QDs, in which relative
energy levels and charging of the two dots can be tuned by an
applied electric field when the QDM is embedded in the Schottky
diode structure~\cite{Bayer01, Krenner05,Stinaff06}. By
time-dependent manipulations with external fields and by varying coupling
parameters quantum gate operations should be realizable. Precise
control over the structure parameters is required for spin
initialization, readout, and coherent manipulation for a practical
application in quantum information processing. However, the number
of parameters describing the QDM in a diode structure is relatively
large~\cite{Loss98, Awschalom02}. These include, the relative
energies of singly and doubly occupied quantum dots, tunneling
amplitudes, effects of spin flips due to spin-orbit coupling and
difference of g-factors~\cite{AG13}.  Optical
studies~\cite{Bayer01, Krenner05} have been used to investigate
selective tunneling of electrons or holes~\cite{Stinaff06, Doty08},
origins of spin fine structure~\cite{Scheibner07} and conditional
quantum dynamics~\cite{Robledo08}.

In this work, we explore the potential of an alternative approach,
called the spin noise spectroscopy (SNS) that has already been 
demonstrated as a powerful probe of physics of hole or electron
spins localized in separated QDs~\cite{Crooker10,Li12}. In this
approach, the spin noise power spectrum is obtained by measuring the
fluctuations of the optical Faraday rotation of a linearly polarized
beam passing through a region with spins. Although up to now, the main
focus of experiments was on the spin noise of an aggregate ensemble
of hundreds of QDs; there is no fundamental limit on SNS to achieve
the level of a single spin sensitivity using Faraday rotation
fluctuations. For example, the Faraday rotation signal of a single
spin has already been demonstrated
experimentally~\cite{single-spin}. Recently, an experiment showed
the possibility to measure spin noise from a single InGaAs QD by 
means of resonant fluorescence~\cite{WarburtonArxiv13}.

Anticipating the future progress with increasing sensitivity of SNS
to resolving single spin fluctuations, in this work, we explore the
potential of this approach for characterization of solid state
nanostructures on a specific example of QDMs. We will show that SNS
may provide several advantages: First, characterization of a QDM can
be achieved at the thermodynamic equilibrium, without applying a
strong perturbation; Second, details of the spin noise power spectrum
can be obtained with high resolution, which can be used not only to
determine resonant frequencies, but also to find the relative integrated
noise power of different noise power peaks, as it was demonstrated
in atomic gases~\cite{Crooker04}.

Moreover, spin noise is derived from the measured Faraday rotation
signal of a light beam. The latter may couple to spins of different
dots with different strengths due to the difference in the frequency
detuning between the measurement beam and the optical resonant
frequencies of two dots. For example, if the measurement is
sensitive only to the spin of one of the QDs, then the tunneling of
an electron or hole will be observed as a change in the measured
signal even if physically the spin does not flip.
We will argue that this effect provides additional means for the
detection of tunneling in the QDM even in the absence of external
magnetic fields.

The structure of this article is as follows. In Sec.~\ref{spinNoise}
we provide basic information about the state of the art of SNS and
the definition of the spin noise power. In Sec.~\ref{SingleCharged}
we consider the model of a QDM charged by a single electron or hole
and propose steps for its full characterization. In
Sec.~\ref{doubleCharged} we investigate possible characteristics of
a more complex case of a QDM with two holes or electrons. We
summarize our results in Sec.~\ref{Concl}.

\section{Spin noise power spectrum}\label{spinNoise}

Spin noise spectroscopy was introduced by Aleksandrov and
Zapasskii~\cite{Aleksandrov81} who employed off-resonant optical
Faraday rotation to passively detect ground-state spin fluctuations
in a gas of sodium atoms. Later it has been refined and applied to
measure various dynamical properties of electron (and hole) spins in
alkali vapors~\cite{Crooker04}, bulk GaAs~\cite{Oestreich05,
Crooker09}, quantum wells~\cite{Muller08} and self-assembled
QDs~\cite{Crooker10}. Dynamic spin fluctuations generate spontaneous
spin precession and decay with the same characteristic energy and
timescales as the macroscopic magnetization of initially polarized
spins. Hence, the physical parameters, such as g-factors, hyperfine
coupling, and spin coherence lifetimes can be determined by
measuring only the correlation spectra~\cite{Oestreich10, Pershin12}.

\begin{figure}
\includegraphics[width=8.0cm]{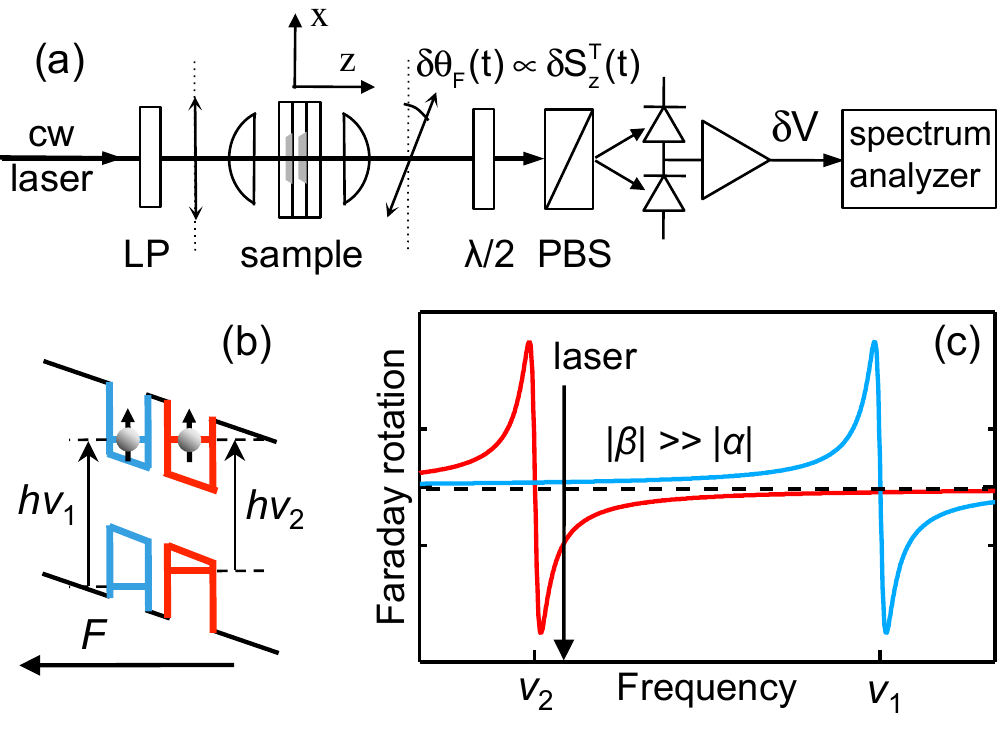}
\caption{(a) Potential experimental setup. LP: linear polarizer; $\lambda$/2: half-wave plate; PBS: polarizing beam splitter. (b) Schematic band
structure of a QDM with two charged electrons in an
applied electric field $F$. (c) Faraday rotation as a function of laser frequency. Due to asymmetries in size, strain and shape, the two QDs have different transition energies. The laser linewidth is much narrower than the absorption linewidth, and close to one QD transition energy (i.e. close to the right quantum dot), so the coupling coefficient $| \beta | \gg | \alpha |$.}
\label{schematic}
\end{figure}

We show in Fig.~\ref{schematic}(a) a potential schematic of the
optical magnetometer to ``listen" to spin noise of a single or double electrons/holes in a QDM. A linearly polarized continuous wave (CW) laser
beam is tuned to have frequency near the resonance of  
transition from the ground-state of the QDM to the higher energy 
level of one of the QDs. Random fluctuations of the electron spins
along the incident beam ($z$-direction), $\delta S_{Lz}(t)$ and
$\delta S_{Rz}(t)$  impart Faraday rotation fluctuations $\delta
\theta_F(t)$ on the transmitted probe-laser beam via spin-dependent
refraction indices for right- and left-circularly polarized light
$n^{\pm}(\nu)$ where $\nu$ is the frequency of the incident beam
\cite{Minhaila06, Crooker09}. Using a half-wave plate and a balanced 
photodiode, these Faraday rotation fluctuations are detected and 
converted to fluctuating voltage signals which are amplified and 
measured by fast digitizers. Power
spectra of the time-domain voltage signals are computed with
fast-Fourier transform algorithms. The Faraday rotation
$\theta_F(\nu)$ scales linearly with $(n^{+}(\nu)-n^{-}(\nu))$, namely, the difference between spin-up and spin down electron densities in the absence of a magnetic field along the direction of laser beam propagation. 

The incident laser beam is detuned sufficiently from the optical
transition energy of both dots to ensure no absorption of the beam
energy by the QDs and the interaction of photons with a QD is mostly
sensitive to the dispersive part of the QD dielectric function.
Importantly, due to asymmetries in the structures of the self-assembled QDs (variations in size, strain, and shape in different QDs), the transition energy from a single resident electron (hole) to negatively (positively) charged trion state can differ from QD to QD [Fig.~\ref{schematic}(b)]. 
The spin noise power spectrum from the QDM is then given by the light intensity
correlation function~\cite{Minhaila06},
 \bea S_N(\omega) =
\int_{0}^{\infty}dt~ \cos {(\omega t)} \la S^T_z(t)S^T_z(0)
\ra,\label{SNPS} \eea
 where we introduced the  {\it weighted spin}:
$S^T_z(t) =\alpha S_{Lz}(t)+\beta S_{Rz}(t)$. Here $\alpha$ and
$\beta$ determine the coupling strengths of the incident laser beam
with the spin in the left and right QDs, and $\la O \ra$ denotes
thermal average of the operator $O$. In the framework of the weak measurement theory,
$\alpha$ and $\beta$ are proportional to the small probabilities per unit of time to collapse the wave function of the QDM and
thus obtain the value of, respectively, $S_{Lz}$ or $S_{Rz}$ at this moment.
The difference $\alpha \ne \beta$ is  due to the fact that 
the detuned laser beam may 
have closer frequency to the resonant frequency of one QD compared to the
other because the detuning in the laser frequency is much smaller
compared to the frequency difference between the resonances of the
two dots [Fig.~\ref{schematic}(c)]. Therefore the coupling of the
incident laser beam with the two QDs is expected to be different,
i.e., $\alpha \ne \beta$~\cite{Zapasskii13}. In fact, the difference
in the detuning energies of two dots is typically as large as a few
meVs, while the Faraday rotation is inversely proportional to the
detuning, which means that generally one should find the regime of
either $\alpha \gg \beta$ or $\alpha \ll \beta$. For completeness,
we will consider an arbitrary ratio $\alpha/ \beta$ in the following
calculations with easily obtainable limiting cases.

In order to predict the spin noise power using
Eq.~(\ref{SNPS}), we calculate eigenvalues and eigenfunctions of
the Hamiltonian of the system, and use them to evaluate the thermal
average of the spin-spin correlation. Let us call $\lambda_i$ and
$|\lambda_i\ra$ eigenvalues and eigenfunctions of the Hamiltonian,
respectively.  Now we write the thermal average of the spin
correlation function as,
\begin{widetext}
\bea
\la S^T_z(t)S^T_z(0) \ra=\sum_{i,j}\f{e^{-\lambda_i/k_BT- |t|/\tau_{ij}}}{\mathcal{Z}}e^{i(\lambda_j-\lambda_i)t  }|\la \lambda_i|S^T_z(0)|\lambda_j\ra|^2, \label{SNPS2}
\eea 
\end{widetext}
where  for each resonance with energy difference $\lambda_j-\lambda_i$, we introduced a phenomenological dephasing time $\tau_{ij}$.
Here the partition function
$\mathcal{Z}=\sum_{i}e^{-\lambda_i/k_BT}$, $k_B$ is the Boltzmann
constant, and $T$ is temperature. 
In the following formulas, in order to simplify notation, we will often suppress decoherence parameters $\tau_{ij}$ in the expression for spin correlators, assuming that they are always there. 
At temperatures $\sim 5$K the
Boltzmann factor can  be disregarded if the tunneling amplitude is
below 1\,GHz (i.e., $\sim1$ $\mu$eV). We will assume such cases of 
weakly coupled QDs.

One can easily suppress tunneling of electrons by applying an electric field $F$ that creates an extra energy difference  $2\epsilon$  between an electron occupying the left or the right dot, $2\epsilon=edF$, where $d$ is the distance between the dots. Relative strengths of parameters $\alpha$ and $\beta$ can be determined simultaneously along with Lande g-factors by  suppressing tunneling  and then applying a magnetic field ${\bf B}$ of 100-1000\,Gauss magnitude. For example, the effective Hamiltonian of a doubly charged QDM in a  magnetic field ${\bf B}\equiv \{b_x,b_y,b_z\}$ can be written as $H_Z=\mu_B{\bf B}\cdot({\bf g}^e_L\cdot{\bf S}_L+{\bf g}^e_R\cdot{\bf S}_R)$. We use ${\bf g}^e_{L} \in \{ g^e_{L\parallel}, g^e_{L\perp}\}$ and ${\bf g}^e_{R} \in \{ g^e_{R\parallel}, g^e_{R\perp}\}$ where $g^e_{L\parallel}$ and $g^e_{L\perp}$ ($g^e_{R\parallel}$ and $g^e_{R\perp}$) are the g-factors of electron in the left (right) QD along the direction ($z$ direction) of the incident linearly polarized laser beam, and the plane perpendicular to $z$ direction respectively, $\mu_B$ is the Bohr magneton. The spin of a single electron or hole inside a QDM makes Larmor precession with a frequency that depends on its position in the QDM. The spin noise correlator in a field $b_x$ at temperatures $\sim 5K$ is given by 
\bea 
\la S^T_z(t)S^T_z(0)\ra &=& \f{1}{2}\big(\alpha^2 e^{-t/T_{2L}}
\cos (\mu_B g^e_{L\perp}b_xt)\nn \\
&+&\beta^2 e^{-t/T_{2R}} \cos(\mu_B g^e_{R\perp}b_xt)\big), 
\eea 
where we take into account phenomenological spin decoherence times $T_{2L}$ and $T_{2R}$ for the left and right quantum dots. When dots are decoupled, such
 decoherence rates can be as small as $1$MHz~\cite{Sinitsyn12}. The
spin noise power spectrum then becomes the sum of two relatively
sharp Lorentzian peaks.
\bea
\nonumber  S_N(\omega)  &=& \f{1}{4}\Big( \frac{\alpha^2 /T_{2L}}{(\omega 
-\mu_Bg^e_{L\perp}b_x)^2+1/T^2_{2L}} +\\ &+&\frac{\beta^2/T_{2R}}{(\omega
-\mu_Bg^e_{R\perp}b_x)^2+1/T^2_{2R}} \Big). \label{lor} \eea Two
finite frequency peaks in the noise power spectrum are centered
around $\mu_Bg^e_{L\perp}b_x$ and $\mu_Bg^e_{R\perp}b_x$. Thus, the
in-plane g-factors of electron in both the QDs can be determined
directly from positions of these peaks. Similarly one can measure
$g^e_{L\parallel}$ and $g^e_{R\parallel}$ by applying an additional
out-of-plane field $b_z$ and measuring the shift of the Larmor
frequencies or changes in the relative areas of power
peaks~\cite{Li12}.

Importantly, as peaks are well separated, one can separately measure
their areas by integrating spin noise power over the regions around
individual peaks. Such an individual peak integrated noise power
does not depend on relaxation rate. For example, the ratio of the
areas of the two peaks in Eq.~(\ref{lor}) would be
$(\alpha/\beta)^2$, which  provides the estimate of the coupling of
the incident laser beam with the electron in either QD.

We study two cases: (1) A QDM is charged
with only a single electron that can hop between two dots, and (2) both dots are charged with an
electron. The Hamiltonians for QDMs containing one excess
hole or two excess holes are qualitatively similar to those of
excess electrons.

\section{Spin noise from single electron charged QDM} \label{SingleCharged}

We consider only the lowest confined energy levels for electrons of each QD of a QDM.
Even if both dots are grown under nominally identical conditions,
the strain and the asymmetry generally lead to non-degenerate dots
with slightly different energy levels. 
The applied electric field $(F)$ controls the relative energy levels
of the two dots at a distance $d$ between the two dots. It can be used to
tune either electron or hole energy levels into the resonance.
Electrons and holes can then tunnel through the barrier separating
the two dots~\cite{Bracker06}. The spatial basis states for the
electrons localized in the left and right QD are given by the two
lowest orthogonal states $\phi_L({\bf r})$ and $\phi_R({\bf r})$. Similarly we define spin basis states $|\sigma,0\ra$ and $|0,\sigma\ra$ for an electron with spin $\sigma$ localized in the left and right QD, respectively. 

When the QDM is charged with a single electron the full
wave-function of QDM is given by $\phi_L({\bf r})|\sigma, 0 \ra$ or $\phi_R({\bf
r})|0,\sigma \ra$ depending on the position of
the electron in the QDM. Electron or hole tunneling between two
quantum dots can be both spin independent and with a spin flipping
with amplitudes $\gamma_c$ and $\gamma_{nc}$, respectively. The
latter is possible due to the spin-orbit coupling. We fix the basis
so that the spin-conserving tunneling $\gamma_c$ is real, and then
we assume that $\gamma_{nc}$ is generally complex. In addition to
the spin-orbit interaction (which we have incorporated in
$\gamma_{nc}$) there are two other sources of spin-dependent local
interaction, one is the Zeeman coupling due to an external magnetic
field and the other is the hyperfine coupling between electron spins
and nuclear spins in a QD. The latter field is time-dependent and
usually is considered the major source of decoherence at low (below
5-10\,K) temperatures~\cite{Li12}. In order to observe coherent
tunneling, this field should be weak so that its other effects could
be disregarded. For example, the Overhauser field for hole doped
InAs QDs is about 25\,Gauss~\cite{Eble09,Li12}. Alternatively, the
decoherence time of 37\,ns in electrostatically defined GaAs quantum
dots~\cite{echo} would also be sufficient for studies of coherent
spin interactions by SNS. Henceforth, we will assume the case of
such QDMs in which the Overhauser field amplitude is negligible in
comparison to other parameters, such as the tunneling amplitude, and
assume that its effect is reduced to introducing the finite
decoherence rate at time scales of 10-100\,ns.

The most general Hamiltonian $H^{(1)}$ of the QDM charged with one
electron, which is allowed by the time-reversal symmetry at zero
magnetic field, in the basis $\{|\uparrow,0\ra,|\downarrow,0\ra,|0,\uparrow\ra,|0,\downarrow\ra\}$
reads
\begin{widetext}
\bea
H^{(1)} = \left( \begin{array}{cccc}\epsilon+g_{L\parallel}b_z &
g_{L\perp}b_- & \gamma_c & \gamma_{nc}\\ g_{L\perp}b_+ &
\epsilon-g_{L\parallel}b_z & -\gamma_{nc}^* & \gamma_c\\\gamma_c &
-\gamma_{nc} & -\epsilon+g_{R\parallel}b_z &
g_{R\perp}b_-\\\gamma_{nc}^* & \gamma_c & g_{R\perp}b_+&
-\epsilon-g_{R\parallel}b_z\end{array}\right),
\label{spHam} \eea
\end{widetext}
where $\epsilon$ is the bias between the two QDs, which can
be controlled by an external electric field;
$g_{L\parallel}=\mu_Bg^e_{L\parallel}/2$,
$g_{R\parallel}=\mu_Bg^e_{R\parallel}/2$,
$g_{L\perp}=\mu_Bg^e_{L\perp}/2$, $g_{R\perp}=\mu_Bg^e_{R\perp}/2$
and $b_{\pm}=b_{x}\pm ib_y$. In (\ref{spHam}), the signs of tunneling
terms are chosen to ensure the time reversal symmetry of the 
spin-orbit coupling.

In the absence of a magnetic field, the eigenvalues of the
Hamiltonian in Eq.~(\ref{spHam}) are doubly degenerate. The
eigenvalues are $\lambda_1=-\lambda \equiv
-\sqrt{\gamma_c^2+|\gamma_{nc}|^2+\epsilon^2},~\lambda_2=-\lambda,
~\lambda_3=\lambda,~\lambda_4=\lambda$, and the corresponding
orthonormal eigenvectors $|\lambda_i\ra$ $(i=1,2,3,4)$ can be
derived explicitly.
We find from Eq.~(\ref{SNPS2}) using $\lambda_i$ and $|\lambda_i\ra$ after disregarding the Boltzmann factor $e^{-\lambda_i/k_B T}$ and suppressing the decoherence parameters $\tau_{ij}$, 
 \bea &&\la
S^T_z(t)S^T_z(0) \ra =
\f{(\alpha^2+\beta^2)(\epsilon^2+\lambda^2)+2\alpha
\beta(\gamma_c^2-|\gamma_{nc}|^2)}{16\lambda^2}\nn\\&&+\f{(\alpha^2+\beta^2)(\gamma_c^2+|\gamma_{nc}|^2)-2\alpha
\beta(\gamma_c^2-|\gamma_{nc}|^2)}{16\lambda^2}\cos (2\lambda
t).\nn\\ \label{SNzf} \eea
The spin noise power spectrum, at
positive frequencies, would correspond to one  peak that is centered
at a frequency $2\lambda$, and one zero-frequency peak. Hence, one
can determine the eigenvalues of the Hamiltonian experimentally from
the position of the finite-frequency peak. In addition, if we
probe the spin noise from the individual QDs of the QDM, we can
calculate both the single-particle energy of the QDs and the total
magnitude of tunneling easily without tuning the electric field, for
example, when $\alpha=0$ and $\beta=1$ (which is a likely case)
the spin noise correlator is given by \bea
\la S_{Rz}(t)S_{Rz}(0) \ra &=& \f{(\epsilon^2+\lambda^2)}{16\lambda^2}+\f{(\gamma_c^2+|\gamma_{nc}|^2)}{16\lambda^2}\cos (2\lambda t).\nn\\
\label{NRQD} \eea
Hence, by measuring the relative integrated powers
of separate peaks, one can determine the bias $\epsilon$ separately
from the effective tunneling parameter
$\sqrt{\gamma_c^2+|\gamma_{nc}^2|}$. In the limit
$\epsilon\gg\gamma_c,|\gamma_{nc}|$, the finite frequency peak in
Eq.~(\ref{NRQD}) vanishes, and $\la S_{Rz}(t)S_{Rz}(0) \ra \to 1/8$.
The meaning of the latter result is the following: When tunneling is
suppressed, the zero-frequency peak becomes merely the average of
the square of the spin operator  $(S_{Rz})^2=1/4$. Since all states
are equally probable, an electron will spend  half of the time on
average in the unobservable QD, which reduces the noise power by an
additional factor of $1/2$.

For self-assembled QDs it is easy to tune the single-particle energy
of the QDs in the QDMs by an applied electric field. In fact, the
applied electric field is also required to charge the QDs with a
single electron in the reverse bias of the Schottky diode
configuration. Therefore we can bring the QDs in a symmetric energy
configuration, in which $\epsilon$ vanishes. In the case of comparable
$\alpha$ and $\beta$, it is possible to derive the values of $\gamma_c$ 
and $|\gamma_{nc}|$ from the relative area (the ratio of the areas) of the zero and finite
frequency peaks, which is
$[(\alpha^2+\beta^2)(\gamma_c^2+|\gamma_{nc}|^2)+2\alpha
\beta(\gamma_c^2-|\gamma_{nc}|^2)]/[(\alpha^2+\beta^2)(\gamma_c^2+|\gamma_{nc}|^2)-2\alpha
\beta(\gamma_c^2-|\gamma_{nc}|^2)]$ with known values of
$\alpha,\beta$. We remind here that the ratio $\alpha/\beta$ can
be found by measuring spin noise at a strong bias in an external
magnetic field.

Alternatively, one can recover all parameters of the Hamiltonian by
applying a weak (in comparison to the tunneling rate) in-plane
magnetic field. To assess the effect of a weak magnetic field we
perform a perturbative calculation.
The degenerate perturbation theory predicts that the lowest order effect of the
magnetic field is a splitting of the otherwise degenerate eigenenergies:
\bea
&&\tilde{\lambda}_{1x\pm}=-\lambda \pm \f{b_x}{2}\Big\{ (g_{L\perp}+g_{R\perp})^2+\f{2\epsilon(-g_{L\perp}^2+g_{R\perp}^2)}{\lambda}\nn\\&&+\f{\epsilon^2(g_{L\perp}-g_{R\perp})^2-4g_{L\perp}g_{R\perp}{\rm Re}[\gamma_{nc}]^2}{\lambda^2}\Big\}^{1/2},\nn\\
&&\tilde{\lambda}_{2x\pm}=+\lambda \pm \f{b_x}{2}\Big\{
(g_{L\perp}+g_{R\perp})^2+\f{2\epsilon(g_{L\perp}^2-g_{R\perp}^2)}{\lambda}\nn\\&&+\f{\epsilon^2(g_{L\perp}-g_{R\perp})^2-4g_{L\perp}g_{R\perp}{\rm
Re}[\gamma_{nc}]^2}{\lambda^2}\Big \}^{1/2}.\nn \eea

Thus, there are two low frequency peaks centered around
$(\tilde{\lambda}_{1x+}-\tilde{\lambda}_{1x-})$ and
$(\tilde{\lambda}_{2x+}-\tilde{\lambda}_{2x-})$ in the spin noise
spectrum for a weak magnetic field $b_x$. A similar perturbative
calculation for an in-plane weak magnetic field $b_y$ provides two
similar low energy peaks whose positions are sensitive to the 
imaginary part of the tunneling parameter $\gamma_{nc}$.

The positions of frequencies at maxima of these four low energy
peaks along with the relative areas of the two peaks at zero
magnetic field in Eq.~(\ref{SNzf}) provide a sufficient number of
algebraic equations from which all the unknown parameters of the
QDM, including relative phases of tunneling rates, can be extracted.
The resolution of the split peaks depends on the width of the peaks
and the relative separation between them. While the width of the
finite-frequency peaks is of the order of 10-50\,MHz, the separation
between the split peaks depends on the applied magnetic field which
can be of the order of 100-1000\,Gauss to resolve them.

\section{Spin noise from double electron charged QDM}\label{doubleCharged}

A doubly charged QDM is the smallest testbed where all the steps
(initialization, readout, and coherent manipulation of spins)
necessary for quantum computation can be realized. In the
spin-blockade regime charge transfer between the two dots of the doubly
charged QDM can be possible only in the spin-singlet sector of two
electrons. This regime has been widely investigated experimentally
to probe the electron and nuclear dynamics. Coherent single spin
manipulations and gate controllable exchange coupling between spins
have been successfully realized in laterally coupled gate-defined
QDs~\cite{Petta05, Koppens06, Nowack07}. Recently coherent optical
initialization, control and readout have also been realized by optical
means in a vertically stacked QDM~\cite{Kim11, Greilich11}.

For a QDM charged by two electrons there are six spin basis
states and four spatial basis states. The spatial parts of
the basis states with single electron occupancy in each QD are given
by symmetric and antisymmetric wavefunctions $\psi_{\pm}({\bf
r}_1,{\bf r}_2)=\f{1}{\sqrt{2}}(\phi_L({\bf r}_1)\phi_R({\bf
r}_2)\pm \phi_R({\bf r}_1)\phi_L({\bf r}_2)).$

The spatial parts of the basis states for the double electron
occupancy in the left and the right QD are respectively
$\psi_{L,R}({\bf r}_1,{\bf r}_2)=\phi_{L,R}({\bf
r}_1)\phi_{L,R}({\bf r}_2)$.
The six basis spin states are three spin-singlets $|\uparrow \downarrow,0\ra_s, |0,\uparrow \downarrow\ra_s, |\uparrow, \downarrow\ra_s$ and three
spin-triplets $|\uparrow, \uparrow\ra_t, |\downarrow, \downarrow\ra_t, |\uparrow, \downarrow\ra_t$. 
The full antisymmetric two electron basis states are $\psi_{L}({\bf r}_1,{\bf r}_2)|\uparrow \downarrow,0\ra_s$, $\psi_{+}({\bf r}_1,{\bf r}_2)|\uparrow,\downarrow\ra_s$, $\psi_{R}({\bf r}_1,{\bf r}_2)|0,\uparrow \downarrow\ra_s$, $\psi_{-}({\bf r}_1,{\bf r}_2)|\uparrow,\uparrow\ra_t$, $\psi_{-}({\bf r}_1,{\bf r}_2)|\uparrow,\downarrow\ra_t$ and $\psi_{-}({\bf r}_1,{\bf r}_2)|\downarrow,\downarrow\ra_t$. The main difference of the two electron case from the single electron charged QDM is the Coulomb interaction $U({\bf r}_1,{\bf r}_2)=e^2/(4\pi \kappa |{\bf r}_1-{\bf r}_2|)$ between two electrons at position ${\bf r}_1,{\bf r}_2$, $\kappa$ is the dielectric constant of the host material. The Coulomb repulsion between two electrons in the left and right dot is given by
$V_{LL}=\int d{\bf r}_1 \int d{\bf r}_2 |\psi_L({\bf r}_1,{\bf r}_2)|^2U({\bf r}_1,{\bf r}_2)$ and $V_{RR}=\int d{\bf r}_1 \int d{\bf r}_2 |\psi_R({\bf r}_1,{\bf r}_2)|^2U({\bf r}_1,{\bf r}_2)$. Because the two dots are not identical, the Coulomb energy between two electrons is different in different dots. The Coulomb interaction between electrons for single electron occupancy in each QD is different for spatially symmetric (corresponds to spin-singlet) and antisymmetric (corresponds to spin-triplet) basis states. They are given by $V^s_{LR}=\int d{\bf r}_1 \int d{\bf r}_2 |\psi_+({\bf r}_1,{\bf r}_2)|^2U({\bf r}_1,{\bf r}_2)$ and $V^t_{LR}=\int d{\bf r}_1 \int d{\bf r}_2 |\psi_-({\bf r}_1,{\bf r}_2)|^2U({\bf r}_1,{\bf
r}_2)$. Thus for $V_{RR}>2|\epsilon|$ two electrons mostly stay as one electron in each QD for a lower energy configuration. However as the energy detuning between the two dots increases, i.e., $V_{RR}<2|\epsilon|$ the ground state becomes $\psi_{R}({\bf r}_1,{\bf r}_2)|0,\uparrow\downarrow\ra_s$. Therefore by changing the applied electric field we can alter the charge (spatial)
configuration of the QDM, which allows us to study the behavior
of spin states in the QDM.

Due to the anisotropy in g-factors of electrons in different QDs, the
Zeeman energy has two parts, one is the total (absolute value) spin-conserving and the
other is the total spin-non-conserving. The Zeeman term in a homogeneous
external magnetic field ${\bf B}$ is written as
$H_Z=\mu_B{\bf B}\cdot({\bf g}^e_L\cdot{\bf S}_L+{\bf g}^e_R\cdot{\bf S}_R)=\f{\mu_B}{2}{\bf B}\cdot({\bf g}^e_L+{\bf g}^e_R)\cdot({\bf S}_L+{\bf S}_R)+\f{\mu_B}{2}{\bf B}\cdot({\bf g}^e_L-{\bf g}^e_R)\cdot({\bf S}_L-{\bf S}_R)$ where the first term in the second expression conserves the magnitude of total spin $|{\bf S}_L+{\bf S}_R|$ of the two electrons, and the second term is total spin-non-conserving.

When the applied magnetic field is homogeneous across the QDs, the
magnitude of total spin-non-conserving Zeeman term depends on the
difference of g-factors. Disregarding the effect of the spin-orbit
interactions, the full Hamiltonian of a two-electron charged QDM in
the spin basis $\{|\uparrow \downarrow,0\ra_s,|\uparrow,
\downarrow\ra_s,|0,\uparrow \downarrow\ra_s,|\uparrow,
\uparrow\ra_t, |\uparrow, \downarrow\ra_t,|\downarrow,
\downarrow\ra_t \}$ then reads~\cite{Doty08, Stepanenko12}
\begin{widetext}
\bea
H^{(2)}=\left( \begin{array}{cccccc} V_{LL}+2\epsilon & -\sqrt{2}\gamma_c & 0 & 0 & 0 & 0 \\-\sqrt{2}\gamma_c & V^{s}_{LR} & -\sqrt{2}\gamma_c & -\sqrt{2}\delta\eta_- & 2\delta\eta_z & \sqrt{2}\delta \eta_+ \\ 0 & -\sqrt{2}\gamma_c & V_{RR}-2\epsilon & 0 & 0 & 0 \\ 0 & -\sqrt{2}\delta \eta_+& 0 & V^{t}_{LR}+2\eta_z & \sqrt{2}\eta_+ & 0 \\ 0 & 2\delta\eta_z & 0 & \sqrt{2}\eta_- & V^{t}_{LR} & \sqrt{2}\eta_+ \\ 0 & \sqrt{2}\delta\eta_- &0 & 0 & \sqrt{2}\eta_- & V^{t}_{LR}-2\eta_z \end{array}\right), \label{mHamMF}
\eea
\end{widetext}
where $\eta_z=(g_{L\parallel}+g_{R\parallel})b_z/2,~\eta_{\pm}\equiv (\eta_x\pm i\eta_y)=(g_{L\perp}+g_{R\perp})(b_x \pm ib_y)/2$ and $\delta\eta_z=(g_{L\parallel}-g_{R\parallel})b_z/2,~\delta\eta_{\pm}=(g_{L\perp}-g_{R\perp})(b_x\pm ib_y)/2$. 
The terms $V^s_{LR}$ and $V^t_{LR}$ account for the Coulomb energy
of the singly occupied singlet and triplet spatial states
$\psi_{\pm}({\bf r}_1,{\bf r}_2)$~\cite{Burkard99}. Here we note
that the parameters $V^s_{LR}$, $V^{t}_{LR}$  and $\gamma_c$ are not
independent in the sense that both $\gamma_c$ and the difference
$V^s_{LR} -V^t_{LR}$ depend on the overlap of the electron wave
functions in different QDs. For example, by increasing the tunneling
barrier, both of them are expected to be suppressed, so the limit
$\gamma_c \rightarrow 0$ should be taken with extra care to account
for similar changes in $V^s_{LR} -V^t_{LR}$. We have not included $\gamma_{nc}$ in Eq. (\ref{mHamMF}) as it makes the Hamiltonian much more complex to study analytically, while not making any significant qualitative changes.

Depending on the value of $\epsilon$ compared to $V_{RR}$ two
electrons can reside in the same dot or in two different dots. In the
absence of a magnetic field, the spin-singlet and the spin-triplet
sectors are decoupled. Thus the spin-conserving part of the
Hamiltonian $H^{(2)}$ reads as \bea H^{(2)}_{sc}=\left(
\begin{array}{cc} H_{SS} & 0\\ 0 & H_{TT}
\end{array}\right),\label{HamSC} \eea where the block $H_{SS}$ is a
$3 \times 3$ matrix in the spin-singlet basis, the block $H_{TT}$ is
a $3 \times 3$ matrix in the spin-triplet basis, and the
off-diagonal blocks connecting the two spin sectors are zero in the
absence of spin-non-conserving interactions.

There is no mixing between basis states in the triplet sectors. In
typical QDMs, $V_{LL},V_{RR} \approx 1$ meV and depend on the
dielectric constant $\kappa$ of the host material. The
spin-conserving tunneling $\gamma_c$ and the Coulomb interaction
strengths $V^s_{LR},V^t_{LR}$ between electrons/holes in different
dots depend on the distance between the dots and the height of the
barrier between them. The accessibility of the spin noise spectroscopy
technique in experiments has been limited to GHz (order of $\mu$eV)
frequency scale so here we mostly concentrate on the regime of a
characteristic tunneling frequency below 1\,GHz. Recently, ultrafast
SNS was introduced that extended the range of applications to
systems with a much faster dynamics, up to hundreds of
GHz~\cite{bersky, Hubner13}. Our results, however, can be easily extended to
other frequency regimes.

\begin{figure}
\includegraphics[width=\columnwidth]{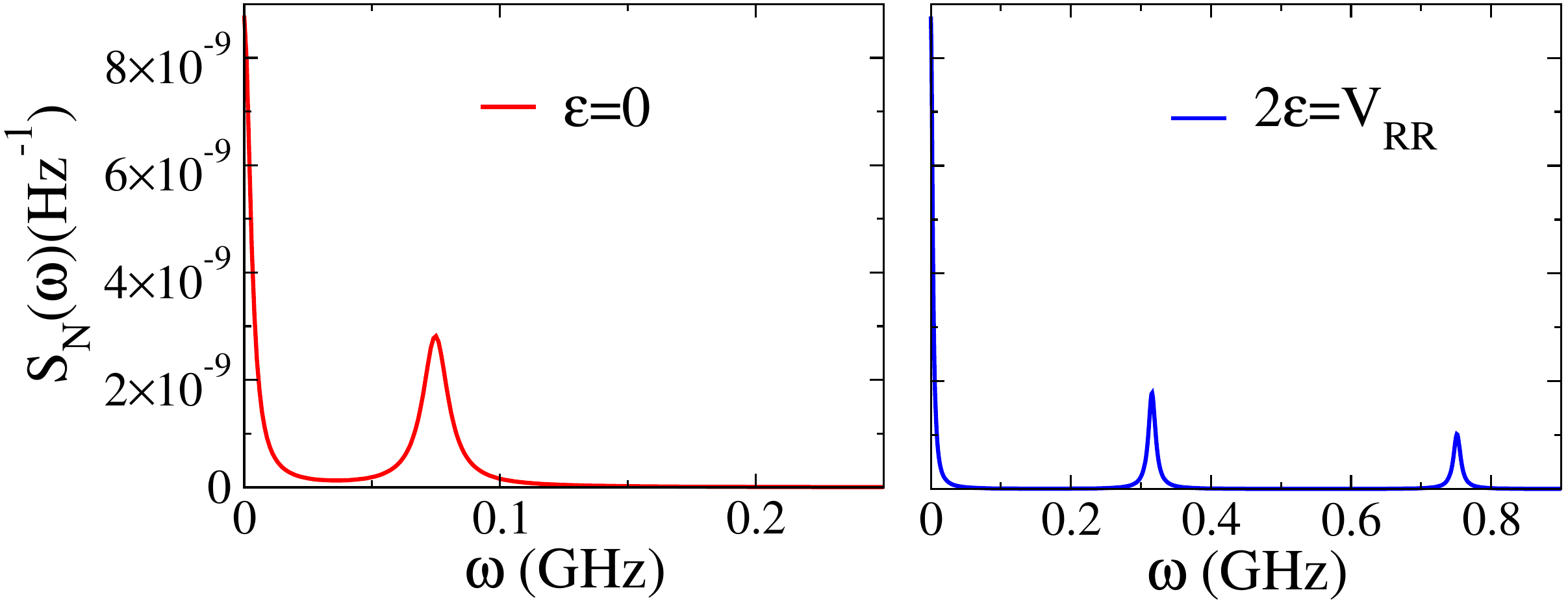}
\caption{Low-frequency spin noise power spectrum for a double
electron charged QDM. The parameters are $V_{RR}=V_{LL}=1$meV,
$\gamma_c=V^{t}_{LR}=1.5\mu$eV, $V^{s}_{LR}=1.2\mu$eV. Broadening
width of the zero frequency and finite frequency peaks are
respectively $0.025\mu$eV and $0.05\mu$eV. The ratio of coupling
$\alpha/\beta=9$.} \label{SN1}
\end{figure}

There can be two types of low frequency/energy resonances in the
doubly charged QDMs. A single finite-frequency peak in the spin
noise spectrum develops when the QDs are almost degenerate in
energy, i.e., $\epsilon \approx 0$. It arises via the virtual
transition of electron between the singlet $|\uparrow,
\downarrow\ra_s$ of single electrons in each QD and the singlets
$|\uparrow \downarrow, 0\ra_s$ , $|0, \uparrow \downarrow\ra_s$ of
two electrons in the same left or right QD, leading to the effective
exchange coupling between spins.  Figure~\ref{SN1}(a) shows that the
spin noise spectrum at $\epsilon \approx 0$ for $V_{LL}\approx
V_{RR}\gg\gamma_c,V^s_{LR},V^{t}_{LR}$ has one zero frequency peak
and one finite frequency peak at
$[(V^s_{LR}-V^{t}_{LR})-2\gamma_c^2/V_{RR}-2\gamma_c^2/V_{LL}]/h$
(in the leading order of $\gamma_c^2/V_{RR}$ and
$\gamma_c^2/V_{LL}$). If the bias $\epsilon$ is set exactly at
resonance, the spin noise correlator reads \bea && \la
S^T_z(t)S^T_z(0)\ra =
\f{(\alpha+\beta)^2}{12}+\f{(\alpha-\beta)^2}{12}\nn\\&& \times 
\cos
\left(V^s_{LR}-V^{t}_{LR}-\f{2\gamma_c^2}{V_{RR}}-\f{2\gamma_c^2}{V_{LL}}\right)t.
\label{SingleReso} 
\eea
The ability to observe (or resolve) the finite frequency peak depends on its width, which is determined mostly by the decoherence rate. The width of the
zero frequency peak is controlled by the spin relaxation time, which
is of the order of 1\,MHz~[\onlinecite{Sinitsyn12}] (see
Fig.~\ref{SN1}), i.e., it is much narrower than the width of the
finite-frequency peak. Two non-zero low frequency peaks can occur
in the spin noise spectrum when the bias $2\epsilon \approx V_{RR}$ [Fig.\ref{SN1} (right panel)].
Then the singlet state $|0,\uparrow \downarrow\ra_s$ of two
electrons in the right QD becomes degenerate with the singlet
$|\uparrow, \downarrow\ra_s$ and triplet $|\uparrow,
\downarrow\ra_t$ states of single electrons in each QD with total
$S_z=0$. In typical QDMs the mixing by spin-conserving tunneling of
one out of three singlet basis states is quite suppressed at a finite
$\epsilon$ because $V_{LL}\approx
V_{RR}\gg\gamma_c,V^s_{LR},V^{t}_{LR}$. For $\epsilon>0$ the singlet
basis state $|\uparrow\downarrow,0\ra_s$ with energy
$V_{LL}+2\epsilon$ is far detuned from the other two singlet basis
states $|\uparrow,\downarrow\ra_s,~|0,\uparrow \downarrow\ra_s$, and
we calculate mixing of the last two basis states. The hybridized
states of the two singlet states are $|S_-\ra=\cos \theta
|\uparrow,\downarrow\ra_s +\sin \theta|0,\uparrow \downarrow\ra_s$
and $|S_+\ra=\sin \theta |\uparrow,\downarrow\ra_s -\cos
\theta|0,\uparrow \downarrow\ra_s$ with energy eigenvalues \bea
E_{S\mp}=\f{V^s_{LR}+V_{RR}-2\epsilon}{2}\mp \sqrt{\f{(V_{RR}-V^s_{LR}-2\epsilon)^2}{4}+2\gamma_c^2},\nn\\
\eea
and $\theta$ is the mixing angle between $|\uparrow,\downarrow\ra_s$ and $|0,\uparrow \downarrow\ra_s$.
When the bias $2\epsilon \approx V_{RR}$ the effective Hamiltonian in the basis
states of $|S_-\ra$, $|S_+\ra$ and $|\uparrow, \downarrow\ra_t$ is a diagonal $3\times3$ matrix with entries,
$E_{s-}|_{2\epsilon \approx V_{RR}}\equiv
\lambda_-=\f{V^{s}_{LR}}{2} -
\sqrt{\f{{V^{s}_{LR}}^2}{4}+2\gamma_c^2}$, $E_{s+}|_{2\epsilon
\approx V_{RR}}\equiv \lambda_+=\f{V^{s}_{LR}}{2}+
\sqrt{\f{{V^{s}_{LR}}^2}{4}+2\gamma_c^2}$ and $V^t_{LR}$
respectively. The spin noise correlators at $2\epsilon \approx
V_{RR}$ are given by \bea
&&\la S^T_z(t)S^T_z(0)\ra = \f{(\alpha+\beta)^2}{12}+\f{(\alpha-\beta)^2}{12}\big(\sin^2\phi  \nn\\&& \times \cos (\lambda_+-V^{t}_{LR})t+  \cos^2\phi  \cos (\lambda_--V^{t}_{LR})t \big), \label{twoReso}\\
&& \sin \phi=\f{-\sqrt{2}\gamma_c}{\sqrt{(V^s_{LR}-\lambda_+)^2+2\gamma_c^2}},\\ &&\cos \phi=\f{(V^s_{LR}-\lambda_+)}{\sqrt{(V^s_{LR}-\lambda_+)^2+2\gamma_c^2}},
\eea
where $\phi=\theta|_{2\epsilon \approx V_{RR}}$. From the relative
area of the zero-frequency and finite-frequency peaks in
Eq.~(\ref{SingleReso}) we  determine the ratio $\alpha/\beta$.
Similarly the relative area of the two finite-frequency peaks in
Eq.~(\ref{twoReso}) can be measured with a good precision, and the
ratio is given by $2\gamma_c^2/\lambda_-^2$. The measured central
frequency of the finite-frequency peak in Eq.~(\ref{SingleReso}) and the
two finite-frequency peaks in Eq.~(\ref{twoReso}), and the relative
area ($\tan^2\phi$) of the two finite-frequency peaks in Eq.~(\ref{twoReso})
provide us four relations for the four unknown parameters $\gamma_c,
V^s_{LR}, V^t_{LR}, V_{RR}$ of the QDM. These relations can be
solved to find the four unknown parameters. Similarly, by tuning
negative $2\epsilon$ near $V_{LL}$ we will have spin noise
correlators as in Eq.~(\ref{twoReso}) with $V_{RR}$ replaced by
$V_{LL}$.

Finally we discuss how the low-frequency spin noise spectrum would
modify in the presence of a homogeneous  total spin-conserving magnetic
field, i.e., $\eta_z,\eta_{\pm}\ne 0$ and
$\delta\eta_z,\delta\eta_{\pm}=0$. We can have $\eta_z,\eta_{\pm}\ne 0$ and $\delta\eta_z,\delta\eta_{\pm}=0$ when ${\bf g}^e_L={\bf g}^e_R$ or ${\bf S}_L={\bf S}_R$. However, we here discuss the spin noise spectrum in a homogeneous magnetic field for same g-factors of the left and right dot electrons, i.e., ${\bf g}^e_L={\bf g}^e_R$. In the presence of $\eta_z,\eta_{\pm}\ne 0$ at a bias $\epsilon \approx 0$, the
finite-frequency peak additionally splits into three peaks with
mean frequencies of two of them shifted by $\pm 2\bar{\eta}$ from the original positions, where the
applied scaled magnetic field
$\bar{\eta}=\sqrt{\eta_x^2+\eta_y^2+\eta_z^2}$. The
peak that was initially centered at the zero frequency splits into a zero-frequency peak and a
finite-frequency peak at a characteristic Larmor frequency $2\bar{\eta}$. Thus there will be  totally four
finite-frequency peaks and a zero frequency peak at a bias $\epsilon
\approx 0$. Similarly we have total seven finite-frequency peaks
and one zero-frequency peak at a bias $2\epsilon \approx V_{RR}$ in
the presence of a homogeneous applied magnetic field.

Here we have not considered the effect of spin-non-conserving
interactions such as the spin-orbit coupling on the spin-noise spectrum
of the doubly charged QDM~\cite{AG13}. The eigenvalues of the
Hamiltonian of the doubly charged QDM in the presence of spin-orbit
coupling would remain doubly degenerate as long as time-reversal
symmetry is preserved. The spin noise spectrum in the presence of
spin-orbit coupling would consequently be similar to the spectrum
which we have derived in the absence of spin-orbit coupling. However
the central frequency of the finite-frequency peaks and the relative
area of the peaks would be renormalized.

\section{Conclusion} \label{Concl}

We discussed the possibility of characterization of quantum dot
molecules by  the spin noise spectroscopy. We showed that this
approach reveals valuable information about parameters of the QDMs,
including coherent tunneling between QDs, without perturbing the
system from the equilibrium.

One of our goals was to emphasize the fact that spin-noise
spectroscopy, when applied to complex nanostructures, is not
restricted to measurements of a total spin dynamics. We predict that
the Faraday effect should be strongly sensitive not only to
magnitude but also to the positions of the spins. This allows one to
resolve coherent tunneling as well as exchange-type spin-spin
interactions even if the total spin is conserved in a process. For
example, the appearance of all finite frequency peaks discussed in
Sec. IV at zero magnetic field would be impossible without the
difference between coupling parameters $\alpha$ and $\beta$. Observations
of such peaks may become an alternative approach to study coherent
tunneling in complex nanostructures.

We explored the spin noise power spectrum of single and double
electron charged QDM in the presence of an applied electric and/or
magnetic field. We argued, in particular, that the relative area of
the noise power peaks contains valuable information for the
characterization of the QDMs. For a single charged QDM, we derived
an explicit expression for the spin noise power in a most general
type of Hamiltonian allowed by the time-reversal symmetry. Both spin
conserving and spin non-conserving tunneling amplitudes (as well as
their relative phase) can be obtained by studying the response of
the noise power spectrum to a weak in-plane magnetic field. A doubly
charged QDM can be explored by similar means. Away from specific
resonances, the spin noise power at finite frequencies will be
suppressed. When the bias electric field is tuned to specific
resonances, the tunneling of electrons/holes between quantum dots
leads to the appearance of new high frequency peaks, whose area and
position can be used to obtain the effective tunneling rate and
other coupling parameters.

We expect that measurements of the noise power spectrum should have
particular advantages over other approaches such as laser absorption
and spin-echo experiments, when characterization of spin coherence
is needed at frequencies below 1\,GHz, e.g. the photoluminescence 
approach is usually resolution limited to several GHz. We believe our 
work will provide a useful guidance for experimental characterization of
quantum nanostructures, in particular QDMs, by means of spin noise spectroscopy.


\section{Acknowledgments}
We thank D. L. Smith and S. A. Crooker for useful discussions. Work at LANL was carried out under the auspices of the Project No. LDRD/20110189ER and the National Nuclear Security Administration of the U.S. Department of Energy at Los Alamos National Laboratory under Contract No. DE-AC52-06NA25396.

\end{document}